\documentclass{ws-ijmpa}
\usepackage{graphicx}
\usepackage{amsmath}
\usepackage[latin1]{inputenc}
\usepackage{amsbsy}
\usepackage{color}
\usepackage{epsfig}
\usepackage{graphicx}
\usepackage{amsmath}
\usepackage{amssymb}
\usepackage{amsbsy}
\usepackage{slashed}
\DeclareMathOperator{\tr}{tr}

\usepackage{xcolor}

\newcommand{\slsh}[1]{{\not \! #1}}

\usepackage[normalem]{ulem}  

\usepackage[colorlinks,citecolor=bl,linkcolor=bl,urlcolor=bl]{hyperref}
\definecolor{bl}{rgb}{.0,.1,.4}

\begin{document}

\title{WEAK MAGNETIC FIELD EFFECTS ON CHIRAL CRITICAL TEMPERATURE IN A NONLOCAL NAMBU--JONA-LASINIO MODEL.}

\author{M. Loewe$^{1,2,3}$, F. Marquez$^1$, C. Villavicencio$^{4}$, R. Zamora$^{1}$}
\address{}
\address{$^1$Instituto de F\'isica, Pontificia Universidad Cat\'olica de Chile, Casilla 306, Santiago 22, Chile}

\address{$^2$Centro Cient\'ifico Tecnol\'ogico de Valpara\'iso, Casilla 110 V, Valpara\'iso, Chile}
\address{$^3$Centre for Theoretical Physics and Mathematical Physics, University of Cape Town, Rondebosch 7700, South Africa}

\address{$^4$Departamento de Ciencias B\'asicas, Universidad del Bio-Bio,\\
 Casilla 447, Chill\'an, Chile}

\maketitle

\begin{abstract}
In this article we study the  nonlocal Nambu--Jona-Lasinio model with a Gaussian regulator in the chiral limit. Finite temperature effects and the presence of a homogeneous magnetic field are considered. The magnetic evolution of the critical temperature for chiral symmetry restoration is then obtained. Here we restrict ourselves to the case of low magnetic field values, being this a complementary discussion to the exisiting analysis in nonlocal models in the strong magnetic field regime.
\end{abstract}

\maketitle

\section{Introduction}

In recent years there has been an increasing interest in studying the QCD phase diagram in the presence of a magnetic field. 
Particularly, the effect of the magnetic field on the critical temperature for chiral phase transition, has been studied in lattice QCD \cite{Braguta,Delia,bali1,bali2,bali3} as well as in different effective models \cite{Boomsma01,Loewe1,Agasian,Fraga,Fraga2,Andersen}. 
Most results from model and lattice calculations have found that magnetic catalysis takes place, i.e. that the critical temperature for chiral phase transition becomes higher in the presence of a magnetic field. However, recent improved lattice calculations have found the opposite behavior \cite{bali1,bali2,bali3}.

The Nambu--Jona-Lasinio (NJL) model and extensions including the Polyakov Loop (PNJL) have been considered for the study of chiral and deconfinement phase transitions in the presence of strong magnetic field  \cite{Gatto:2010pt,Boomsma01,Gatto01,Hell:2009by,Hell:2011ic,Kashiwa:2011td,Horvatic:2010md,Radzhabov:2010dd,Ferreira:2013tba}.

The nonlocal Nambu--Jona-Lasinio (nNJL) models are an attempt to improve NJL model in a more realistic way, inspired in low energy approaches as Dyson-Schwinger ressumation, lattice results, instantons liquid model and one gluon exchange models \cite{Frasca01,Scoccola05,Weise01,Alford01,Alford02}. 
The use of a gaussian regulator in a nNJL model goes back to \cite{Bowler:1994ir,Schmidt:1994di,Golli:1998rf}. 
In this context, the external magnetic field effects on the critical temperature for chiral restoration and deconfinement was studied in  \cite{Kashiwa01} for the case of a strong magnetic field, $eB>10m_\pi^2$, where the approximation used was to cut the Landau series.
For the weak magnetic field case it is necessary to sum over too many Landau levels in order to obtain an accurate result.
Another approach is to expand the fermion propagator in powers of $eB$, which is possible if the magnetic field is smaller than the square of the lowest particle energy, in this case, the lowest Matsubara frequency, i.e.  $eB<(\pi T)^2$ \cite{Ayala03,Ayala04}.
In this article we consider magnetic effects in the nNJL model with a Gaussian regulator, in the regime of low magnetic field, in order to compare with previous discussions in the strong field case \cite{Kashiwa01}. This will give also a better understanding on the validity of the expansion of the fermion propagator in powers of $eB$.
To study this we restrict ourselves to the chiral limit, in which case the chiral phase transition at vanishing chemical potential is a second order one \cite{Scoccola04}, and the result will be compared with NJL with and without the weak magnetic field expansion.\\

This article is organized as follows: In Sec. II the NJL and nNJL models are introduced and the appropriate gap equations are computed. In Sec. III the effect of the magnetic field on the critical temperature for the chiral phase transition is shown. 

Finally, in Sec. IV, we present  our conclusions.\\
\section{The model}

The NJL model and its nonlocal variant (nNJL) have been vastly used to study the thermodynamics of the low energy limit of QCD (see e.g. \cite{Buballa01,Klevansky01,Blaschke01,Scoccola01} and references therein). Both models have an approximate chiral symmetry and present a chiral phase transition. The Lagrangian for the NJL model is
\begin{equation}\mathcal{L}=\bar{\psi}(x)(i\slashed{\partial}-m)\psi(x)+\frac{G}{2}\left(\left(\bar{\psi}\psi\right)^2+\left(\bar{\psi}i\gamma^5\boldsymbol{\tau}\psi\right)^2\right)\label{lagrangiannjl},\end{equation}
with $\boldsymbol{\tau}$ the Pauli matrices in isospin space and $\psi(x)$ a quark field. In the mean field approximation, the quarks acquire an effective mass $M=m+G\langle\bar{\psi}\psi\rangle$. The dressed propagator can be written in Euclidean space as
\begin{equation}
S_E(q)=\frac{-\slashed{q}+M}{q^2+M^2}.
\end{equation}
The value of the constituent mass $M$ can be determined through the {\em gap equation}, obtained by minimizing the effective potential with respect to the mean field \cite{Buballa01,Klevansky01}
\begin{equation}
M-m=G\int\frac{d^4q}{(2\pi)^4}\tr(S_E(q)),\label{gap1}
\end{equation}
where the trace goes over color, flavor and Lorentz indices. The NJL model is nonrenormalizable and the integral in the previous equation needs to be regularized. This can be done in different ways. Since we are interested in studying the model coupled to a magnetic field, the proper time regularization method turns out to be appropriate \cite{Schwinger01}.  Inserting
\begin{equation}
S_E(q)= (M-\slashed{q}) 
\int_\eta^{\infty}{ds}
e^{-s(M^2+q^2)},
\end{equation}
where $\eta=1/\Lambda^2$ is an UV cutoff, in Eq. (\ref{gap1}) and performing the momentum integrals, the gap equation now reads
\begin{equation}M-m=MG\frac{N_cN_f}{4\pi^2}\int_\eta^{\infty}\frac{ds}{s^2}{\rm e}^{-s M^2}.\label{gap2}\end{equation}
Following the prescription from \cite{LeBellac}, a finite temperature gap equation can be obtained by considering the Matsubara frequencies $\omega_n$, such that
\begin{eqnarray}
q_4&\rightarrow& \omega_n=(2n+1)\pi T\label{FT1}\\
\int\frac{dq_4}{2\pi}&\rightarrow& T\sum_{n=-\infty}^{\infty},\label{FT2}
\end{eqnarray}
in Eq. (\ref{gap1}). This yields
\begin{equation}M-m=M\,G\,T\frac{N_cN_f}{2\pi^{3/2}}\sum_n\int_\eta^{\infty}\frac{ds}{s^{3/2}}{\rm e}^{-s(\omega_n^2+M^2)},\label{gap3}\end{equation}
where $N_f=2$ and $N_c=3$. We are interested in studying the model coupled to a homogeneous magnetic field. The derivative in the Lagrangian (\ref{lagrangiannjl}) is replaced by a covariant derivative
\begin{equation}
D_\mu=\partial_\mu + iq_fA_\mu.
\end{equation}
where $A^{\mu}$ is the vector potential corresponding to an homogeneous external magnetic field $\boldsymbol{B}=|\boldsymbol{B}|\hat{z}$ and $q_f$ is the electric charge of the quark fields (i.e. $q_u = 2e/3$ and $q_d = -e/3$). In the symmetric gauge,
\begin{equation}
A^{\mu}= \frac{B}{2}(0,-y,x,0),
\end{equation}
The Schwinger proper time representation for the propagator is given by \cite{Schwinger01}
\begin{multline}
S_E(p)=\int_{\eta}^\infty ds \frac{e^{-s(p_{\|}^2+p_\perp^2
   \frac{\tanh (qBs)}{qBs} + M^2)}}{\cosh (qBs)}
   \\\times\biggl[\left(\cosh (qBs) -i \gamma_1 \gamma_2 \sinh (qBs)\right)
   (M-\slsh{p_{\|}}) - \frac{\slsh{p_\bot}}{\cosh(qBs)} \biggr],
\end{multline} 
with $p_\|^2=p_4^2+p_3^2$, $p_\bot^2=p_1^2+p_2^2$ and where $q$ is the charge of the particle being $B$ the magnetic field. 
Strictly speaking, the Green function involves the presence of a nonlocal phase.
However, since we are dealing  with a  closed one-loop diagram, this phase does not contribute and, therefore,  it will be not considered in what follows.

Using this propagator, we can obtain a zero temperature gap equation in the presence of a magnetic field
\begin{equation}
M-m=GM\frac{N_c}{4\pi^2}\sum_{f=u,d}|q_fB|\int_\eta^{\infty}\frac{ds}{s\tanh(|q_fB|s)}{\rm e}^{-s M^2}.\label{gap4}
\end{equation}
with $q_u=2e/3$ and $q_d=-e/3$. Similarly, the finite temperature gap equation reads
\begin{equation}
M-m=GM\frac{N_c}{\pi^{3/2}}T\sum_{f=u,d}|q_fB|\sum_n\int_\eta^{\infty}\frac{ds\;{\rm e}^{-s (\omega_n^2+M^2)}}{\sqrt{s}\tanh(|q_fB|s)}.\label{gap5}
\end{equation}
Frequently, when working with  NJL-type  models,  the bozonization procedure is incorporated, identifying  the bosonic fields as $\sigma(x)=G\bar{\psi}(x)\psi(x)$ and $\boldsymbol{\pi}(x)=G(\bar{\psi}(x)i\gamma^5\boldsymbol{\tau}\psi(x))$ (see \refcite{Scoccola03} for more details). 
In the mean field approximation, i.e. $\sigma\approx\bar{\sigma}=G\langle\bar{\psi}\psi\rangle$ and $\boldsymbol{\pi}\approx\bar{\boldsymbol{\pi}}=0$, the mean field value of the $\boldsymbol{\pi}$-fields vanishes due to parity conservation and isospin symmetry.
In this way $\bar{\sigma}=M-m$ and, since the $\bar{\sigma}$ field is related to the chiral condensate, the temperature at which $\bar{\sigma}(T)=0$ corresponds to the critical temperature for chiral phase transition, if $m=0$ (chiral limit).
 Equations (\ref{gap2}), (\ref{gap3}), (\ref{gap4}) and (\ref{gap5}) allow us to get $\bar{\sigma}(T, qB)$ which in turn provides us with a magnetic field dependent  critical temperature  $T_c(qB)$ for chiral phase transition.\\

The Lagrangian for the nNJL model is
\begin{equation}\mathcal{L}=\bar{\psi}(x)(i\slashed{\partial}-m)\psi(x)+\frac{G}{2}j_a(x)j_a(x),\end{equation}
with the nonlocal currents
\begin{equation}j_a(x)=\int d^4y\,d^4z r(y-x)r(x-z)\bar{\psi}(y)\Gamma_a\psi(z)\end{equation}
and where $\Gamma_a=(1, i\gamma^5\boldsymbol{\tau})$. The function $r$ in the previous equation is called the {\em regulator} of the model. The regulator may take different forms \cite{Praszalowicz01,Scoccola02}, ``e.g"
the  instanton liquid model suggests in a natural way a Gaussian regulator \cite{Weise01}.
The usual bosonization procedure can be performed similarly to what we did in the NJL model. 
By taking the mean field approximation with a vanishing value for the vacuum expectation value of the pionic fields, the dressed propagator of the model can be written as
\begin{equation}
S_E(q)=\frac{-\slashed{q}+\Sigma(q)}{q^2+\Sigma^2(q)}.\label{nNJLprop}
\end{equation}
As seen in Eq. (\ref{nNJLprop}), the constituent mass of the NJL model has now been replaced by $\Sigma(q)=m+\bar{\sigma}r^2(q)$. 
As in the NJL model, the temperature for which $\bar{\sigma}(T)=0$ is the critical temperature for the chiral phase transition. In order to get the thermal evolution of $\bar{\sigma}$ one needs to solve the gap equation. The zero temperature gap equation reads
\begin{equation}\bar{\sigma}=G\frac{N_c}{\pi^2}\int_0^\infty dp\, p^3\frac{r^2(p)\Sigma(p^2)}{p^2+\Sigma^2(p)}.\label{gap6}\end{equation}
The momentum integrals can now be computed since the nNJL model does not need a UV cutoff. However, an energy scale $\Lambda$ is hidden within the regulator $r(p)$. We can obtain the corresponding finite temperature gap equation following the prescriptions given in Eqs. (\ref{FT1}) and (\ref{FT2})
\begin{equation}\bar{\sigma}=G\, T\frac{4N_c}{\pi^2}\sum_n\int_0^\infty dp\, p^2\frac{r^2(\omega_n,p)\Sigma(\omega_n,p)}{\omega_n^2+p^2+\Sigma^2(\omega_n,p)}.\label{gap7}\end{equation}
Once again, we will resort to the Schwinger representation of the propagator in order to obtain the gap equations in the presence of a homogeneous magnetic field. 
A natural extension for 
the zero temperature gap equation  in the presence of a magnetic field reads
\begin{equation}\bar{\sigma}=G\frac{N_c}{4\pi^4}\sum_{f=u,d}\int_0^\infty ds\int_0^\infty d^4p \,r^2(p)\Sigma(p){\rm e}^{-s\left(\Sigma^2(p)+p_\|^2+\frac{\tanh(|q_fB|s)}{|q_fB|s}p_{\bot}^2\right)}.
\label{gap8}
\end{equation}

Certainly this is an approximation based on the simple repacement of the constant mass by a  running mass $M\to\Sigma(\omega_n,\bm{q})$.
Similar kind of replacements are common in the literature,  for example when  the perpendicular momentum is replaced by the lowest Landau level.\cite{Kashiwa01}

The finite temperature gap equation is obtained from Eq. (\ref{gap7}) by using the prescriptions in Eqs. (\ref{FT1}) and (\ref{FT2}).

\section{Results}

The gap equations introduced in the previous section will allow us to compute $\bar{\sigma}(T, qB)$ both for the NJL and nNJL models. We can then look for the critical temperature for the chiral phase transition obtaining its dependence on the magnetic field, i.e. $T_c(qB)$. Since we are interested in studying how the magnetic field affects the critical temperature for the chiral phase transition, we will work in the chiral limit $m=0$. If we do not do this, then the transition between the broken and restored phases is rather a crossover, which makes the definition of a critical temperature a bit ambiguous. Having fixed the value of the current mass $m$ to zero, the NJL model still has two parameters that need to be fixed, namely $\Lambda$ and the coupling constant $G$. We take $\Lambda=1086$ MeV and $G\Lambda^2=7.56$ from \cite{Klevansky01}, being then our energy scale $\Lambda$ much bigger than the temperatures and magnetic fields involved in our analysis. Using Eq. (\ref{gap2}) we can also determine $M(T=qB=0)=\bar{\sigma}(T=qB=0)=200$ MeV. We can then use the gap equation in the presence of a magnetic field to get $T_c(qB)$.\\

\begin{figure}[!htb]
\begin{center}
\includegraphics[scale=0.45]{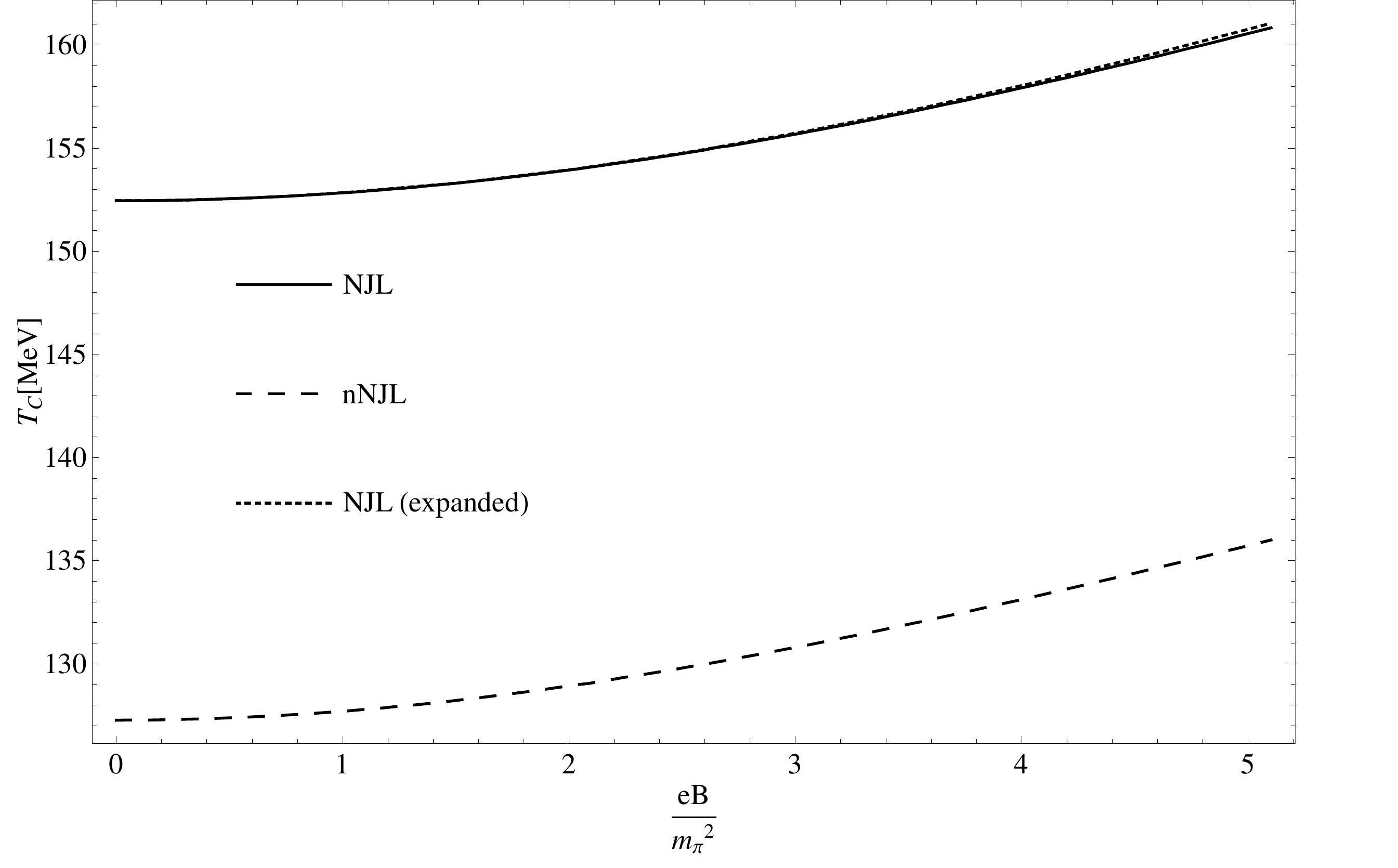}
\caption{Critical temperature for the chiral phase transition as a function of $eB$ for the NJL and nNJL models.
The NJL model is presnted for two cases:  the solid line (full propagator) and the dotted line (magnetic field expansion).
The dashed line represent the nNJL model expanded in powes of $eB$
}
\label{fig1}
\end{center}
\end{figure}

\begin{figure}[!htb]
\begin{center}
\includegraphics[scale=0.5]{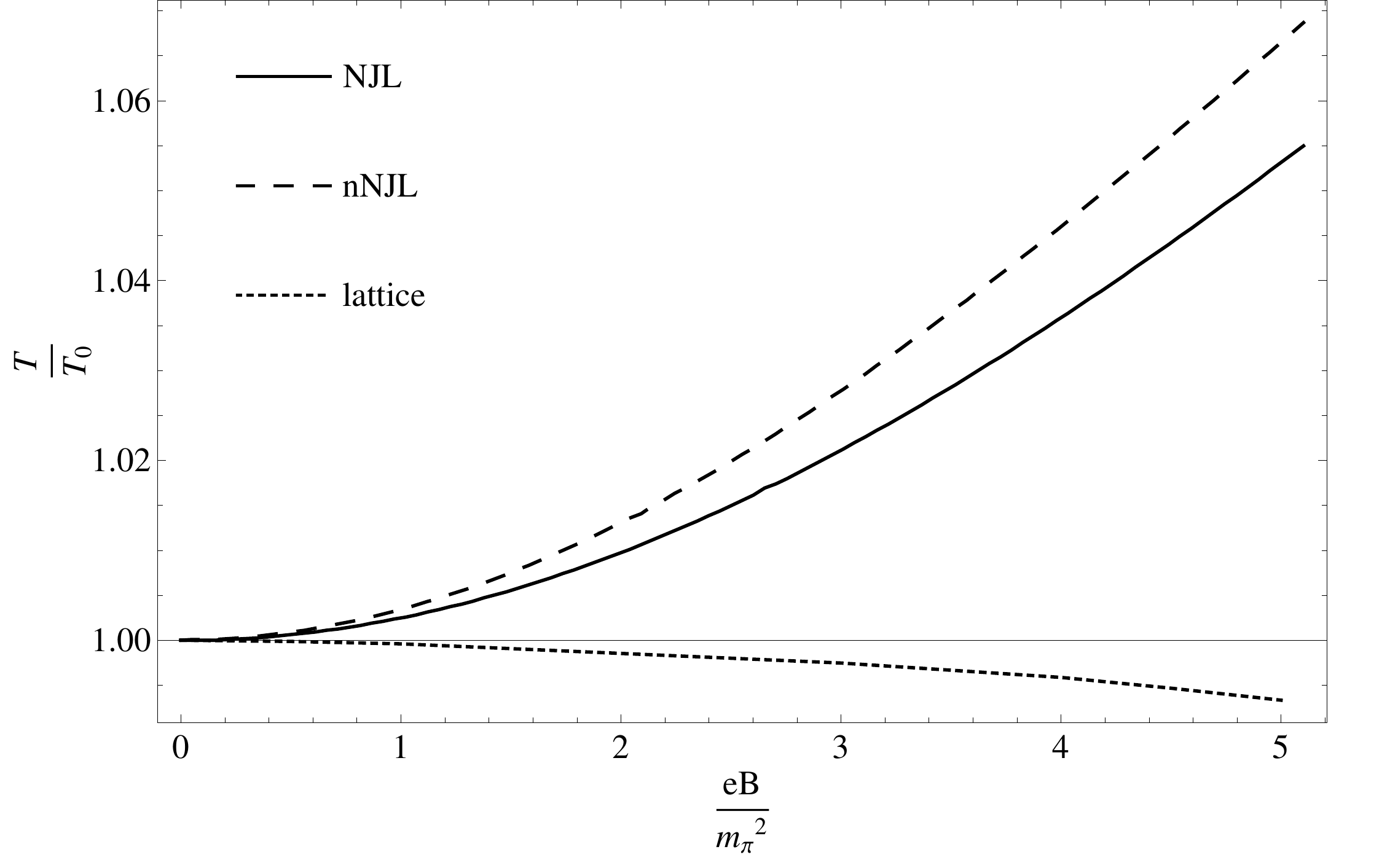}
\caption{Chiral critical temperature, normalized   by its value at $B=0$ for each model, as function of $eB$. 
The dashed line corresponds to the nNJL model scaled with $T_0= 127$~MeV.
The solid line corresponds to the NJL model scaled with $T_0=153$~MeV. 
The dotted line shows recent lattice results scaled with $T_0=150$~MeV.} 
\label{fig2}
\end{center}
\end{figure}

As seen in Fig. \ref{fig1} and \ref{fig2} the critical temperature for the chiral phase transition rises with $qB$. For the case $B=0$, the critical temperature was found around $T\approx152$ MeV. This is known as magnetic catalysis. In order to test the validity of the expansion in powers of $eB$ in the weak field case, we solved the NJL model in this limit finding similar results to those of the exact calculation.\\

For the nNJL model we will consider a Gaussian regulator inspired in the instanton liquid model \cite{Frasca01,Weise01}
\begin{equation}r^2(q_E^2)={\rm e}^{-q_E^2/\Lambda^2}.\end{equation}
We fix the free parameters of the model in the chiral limit using as inputs the pion decay constant $f_\pi=90$~MeV and the chiral condensate $\langle\bar{q}q\rangle=-(260\textrm{MeV})^3$.\cite{Scoccola03} 
This yields $\Lambda=914$ MeV and $G=21\cdot10^{-6}$ MeV${}^{-2}$. 
From the gap equation (\ref{gap6}) we also get $\bar{\sigma}(T=qB=0)=235$ MeV. 

For simplicity, we will consider the weak magnetic field case,  expanding in powers of $eB$ up to order $(eB)^2$. 
Such approximation is valid at finite temperature whenever $|eB|<(\pi T)^2.$\cite{Ayala04}
From Eq. (\ref{gap7}) we can determine the critical temperature for the chiral phase transition in the absence of a magnetic field to be $T_c\approx127$ MeV. 
Then we will consider values for the magnetic field according to above mentioned restrictions.
The fermionic propagator in this region can be written as \cite{Chyi}
\begin{eqnarray}
&&S_E(p) = \frac{(\Sigma -\slsh{p})}{K^2+m_f^2} - i\frac{\gamma_1 \gamma_2(qB) (\Sigma -\slsh{p}_{\|})}{(p^2+\Sigma^2)^2} \nonumber \\
&+&\frac{2 (qB)^2 p_{\bot}^2}{(p^2+\Sigma^2)^4} \biggl[ (\Sigma-\slsh{p}_{\|}) + \frac{\slsh{p}_{\bot}(\Sigma^2+{p}_{\|}^2)}{p_{\bot}^2} \biggr].
\end{eqnarray}
In this case, the finite temperature gap equation in the pressence of an homogeneous magentic field is
\begin{multline}
\bar{\sigma}=G\,T\,\frac{2N_c}{\pi^2}\sum_{f=u,d}\sum_n\int dp\, p^2\,r^2(\omega_n, p)\\\times\left[\frac{\Sigma(\omega_n,p)}{p^2+\omega_n^2+\Sigma^2(\omega_n,p)}+\frac{4}{3}\frac{|q_fB|^2\left(p^2+\omega_n^2\right)\Sigma^2(\omega_n,p)}{\left(p^2+\omega_n^2+\Sigma^2(\omega_n,p)\right)^4}\right].
\end{multline}

The efficacy of the expansion in powers of $eB$, can be seen in Fig. \ref{fig1} where we compare the critical temperature for the expanded and non expanded NJL model, together with the expanded nNJL.
For the region proposed, the expansion of NJL gives us  basically the same results than using the full propagator. 
We can expect the same situation for the full nNJL model and its expanded version also showed in this figure.
Since the magnetic field is usually written in units of $m_\pi^2$,  we adopt this scaling with $m_\pi=140$~MeV.
Notice that this expansion allows us to reach  magnetic field values up to $\sim 5m_\pi^2$, which is bigger than the expected values in magnetars and in the range of the expected generated  fields in perpherial heavy ion collisions.

In Fig. \ref{fig2} we show also the critical temperature for chiral phase transition as function of the magnetic field.
In this case, however,  we compare the two models considered in this paper, the NJL model and the expanded nNJL, with recent lattice results.\cite{bali1,bali2,bali3}
Each line is normalized by their respective critical temperature at zero magnetic field: $T_0^{\mathrm{(NJL)}}=153$~MeV,  $T_0^{\mathrm{(nNJL)}}=127$~MeV, $T_0^{\mathrm{(lattice)}}=150$~MeV.
We can see that the percentual increase in the critical potential is higher in the (expanded) nNJL model, than the NJL model.
It can be seen that the slopes at the origin in both models vanish since they behave quadratically as a function of $B$.
Here we try to explore this nonlocal model with the minimal ingredients, and therefore, the disagreement with recent lattice results is not surprising. 
In fact, some proposals have been made trying to explain this mismatch with lattice results, for example by incorporating extra interactions \cite{Ferreira:2013tba} or by taking into account thermo-magnetic corrections to the coupling constants and screening effects \cite{Ayala1,Ayala2}. 
Theese considerations suggest an appropriate extension of the nNJL model could provide a better agreement.

\section{Conclusions}

We have studied the NJL and nNJL model in the presence of a homogeneous magnetic field, where  in the nNJL model we use a Gaussian regulator and the magnetic effects  are incorporated through  an appropriate extension of  the Schwinger propagator. 
The chiral critical temperature is calculated for the NJL model, and for the nNJL expanded up to second order in the magnetic field ($eB<5m_\pi^2$). We show that in the NJL this expansion gives us the same result than using the full propagator (we expect the same will happen in the  nNJL model). 
In both cases we found that the critical temperature for the chiral phase transition increases as the magnetic field grows, which is consistent with the catalysis found in most of the effective models discused in the literature.
We find that the increase of the critical temperature due to magnetic field effects is percental bigger in the nonlocal model. 


It could be worthwhile to consider scenarios where the discussion of magnetic effects  in nNJL is extended by including  corrections beyond mean field. We leave those extensions for future work.

\section{Acknowledgements}

ML and CV  would like to thank support from FONDECYT under grant No. 1130056, 1150847 and 1150471. FM acknowledges support from CONICYT under grant No. 21110577 and RZ acknowledges support from CONICYT under grant 21110295.

\bibliography{NJL}{}
\bibliographystyle{ws-ijmpa}

\end{document}